# LEQA: Latency Estimation for a Quantum Algorithm Mapped to a Quantum Circuit Fabric

Mohammad Javad Dousti and Massoud Pedram
Department of Electrical Engineering, University of Southern California, Los Angeles, CA 90089, U.S.A.
{dousti, pedram}@usc.edu

## ABSTRACT
This paper presents *LEQA*, a fast latency estimation tool for evaluating the performance of a quantum algorithm mapped to a quantum fabric. The actual quantum algorithm latency can be computed by performing detailed scheduling, placement and routing of the quantum instructions and qubits in a quantum operation dependency graph on a quantum circuit fabric. This is, however, a very expensive proposition that requires large amounts of processing time. Instead, LEQA, which is based on computing the neighborhood population counts of qubits, can produce estimates of the circuit latency with good accuracy (i.e., an average of less than 3% error) with up to two orders of magnitude speedup for mid-size benchmarks. This speedup is expected to increase superlinearly as a function of circuit size (operation count).

## Categories and Subject Descriptors
B.7.2 [**Integrated Circuits**]: Design Aids – *Simulation, Placement and routing*.

## 1. INTRODUCTION
To accurately calculate the latency (total execution time) of a software program, one needs to simulate or run it on a *specific* processor. Changing the processor architecture including the size of cache memories or internal buffers can affect the latency dramatically. A number of approaches have been proposed to estimate program latency without performing time consuming simulations [1][2]. Researchers in the area of quantum computing face the same issue for estimating the latency of a quantum algorithm, programmed in a high-level quantum programming language such as QPL. In this field, the problem is even harder because the size of quantum programs for real-size problems is so huge that the simulation time is much more time consuming than that for classical programs [3].[1]

Devising a new quantum algorithm is a challenging task because of the complex structure of today's quantum computers and the non-intuitive principles (i.e. quantum physics) they are built upon. Currently, quantum algorithms are designed and evaluated by asymptotic runtime analysis, i.e. big $\mathcal{O}$ notation [4]. Unfortunately, in many cases, the asymptotic analysis is too coarse-grained to be of practical use to algorithm developers. Another problem is that quantum computers built using the current technology are only capable of executing toy-size programs, so they cannot be used to experimentally determine the latency of a quantum program. Hence devising a fast, yet accurate, method for estimating the latency of a program is necessary. This method would enable quantum algorithm designers to evaluate their new algorithms and *learn* efficient ways of *coding their quantum algorithms* by quickly comparing the latency of different software coding techniques. Moreover, this method allows designers of *quantum error correction codes* (QECC) to investigate the effect of different error correction codes on the latency of quantum programs.

Latency is an important factor for QECC designers since quantum computers allow only a limited amount of time for running a quantum program without using error correction. QECC has a high impact on the latency. At the same time, one needs to know the latency of a quantum program to know how much error correction it needs. So there is a complex inter-dependency between the quantum algorithm and its latency on one hand and the QECC used on the other hand.

In this paper, we present a procedural method to accurately and quickly estimate the latency of a quantum program. A tool called *quantum algorithm latency estimator* (LEQA) is developed based on this method. To the best of our knowledge, no research has been conducted on this topic before.

The rest of this paper is organized as follows. Section 2 uses the prior art (such as [5] and [6]) to describe a (somewhat novel) design flow for compiling a quantum algorithm and mapping it to primitive quantum structures on a 2-D plane. Section 3 explains the estimation method used for the latency calculation. A procedural method is presented for estimating the average routing latency for the CNOT gates. This section introduces a new parameter called $d_{uncong}$, which is the average routing latency of a qubit in an average-size presence zone when the routing channels are not congested. Estimation of this parameter is explained followed by the detailed description of LEQA (a prototype software implementation of the proposed method). Section 4 presents experimental results while Section 5 concludes the paper.

## 2. A QUANTUM DESIGN FLOW
A typical quantum circuit fabric consists of an infinite 2-D array of identical primitive structures (called *quantum templates* in this paper), each structure containing some sites for generating/initializing qubits, measuring them, performing operations on one or two qubits, and channels for moving qubits or swapping their information. Unfortunately, dealing with this primitive template array is very cumbersome and unwieldy. So in practice another 2-D array of super-templates (which we call *tiles*) is built. Each tile comprises a number of primitive templates. Instead of mapping a quantum circuit directly to the quantum fabric, quantum circuit is mapped to this tiled architecture (see below). A quantum logic synthesis tool (surveyed in reference [7]) generates a reversible quantum circuit. Every qubit in the output circuit is called a *logical qubit*, which is subsequently encoded into several *physical qubits* to detect and correct potential errors.

---
[1] By simulation, we only mean tracing the execution of quantum operations. Simulation of a quantum program and calculating the results cannot be performed efficiently on classical computers even for mid-size problems.

To prevent the propagation of errors in the quantum circuit, the (reversible) logic gates in the synthesized circuit (which are typically NOT, CNOT, and Toffoli gates [8]) must be converted into *Fault-Tolerant* (FT) quantum operations. A possible universal (but redundant) set of FT quantum operations includes CNOT, H (Hadamard), T ($\pi/4$ rotation), $T^\dagger$ (-$\pi/4$ rotation), S (phase), X, Y and Z gates. Note that these gates are all one and two-qubit gates. Implementation of these FT quantum operations depend on the picked error correction method. Note that the set {CNOT, H, T} constitutes a universal basis for quantum circuit realization–the other operations are included to enable more logical simplification in the process of converting the logic synthesis output to the FT quantum operation realization. Each quantum fabric is natively capable of performing a universal set of one and two-qubit instructions (also called native quantum instructions). This set differs among various quantum fabrics. Each FT quantum operation can be implemented by using a composition of these native quantum instructions.

The transformation from logical gates (results of the quantum logic synthesis) to the FT quantum operations and from the FT quantum operations to the native quantum instructions can be called *quantum FT synthesis* and *quantum fabric synthesis*, respectively. Quantum FT and quantum fabric synthesis are outside of the scope of this paper. Each of these FT quantum operations performs a desired function on one or two logical qubits as input producing one or two logical qubits as output; each of the input qubits is encoded with some number of physical qubits. The output qubits will also be coded. Moreover, each of these FT quantum operations requires syndrome extraction circuitry following the quantum gate in order to detect and correct errors (up to a certain limit) that may have been introduced by the quantum operation. Based on the adopted encoding scheme, implementation of each of the aforementioned FT quantum operations may require hundreds to tens of thousands of native quantum instructions in a given quantum fabric.

Various works (e.g. [9] and [10]) have suggested using the *tiled quantum architecture* (TQA), composed of a regular two-dimensional array of *Universal Logic Blocks* (ULBs) to avoid dealing with this complexity. Notice that each ULB in TQA is capable of performing any FT quantum operations. ULBs are separated by the routing channels, which are needed to move logical qubits (or information about these qubits) from some source ULBs to a target ULB in the TQA. A pictorial representation of the TQA is shown in Figure 1. The quantum structures placed at the junctions of routing channels may be thought out as *quantum crossbars* (possibly with some qubit purification capability [11]). Routing channels and quantum crossbars are also built from quantum templates.

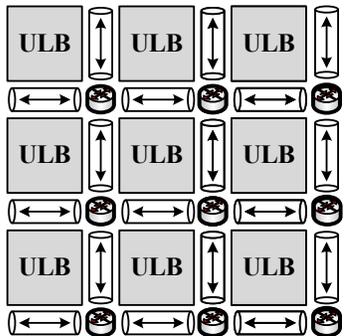

**Figure 1. A 3×3 tiled quantum architecture (TQA)**

A ULB is analogous to a *Configurable Logic Block* (CLB) in an FPGA device, in that it can implement any of a set of target functions. Moreover, the same ULB (as identified by its unique row and column indices in the ULB array) can be configured to perform different FT quantum operations at different times as needed. This is analogous to an on-the-fly-reconfigurable CLB. After appropriate high-level transformations, a quantum algorithm may be represented as a *quantum operation dependency graph* (QODG), in which nodes represent FT quantum operations and edges capture data dependencies. A one-qubit operation is represented by a node with one edge entering it and one edge leaving it. On the other hand, a two-qubit operation is shown using a node with two edges entering it and two edges leaving it. One edge is called *control edge* while the other is called *target edge*. It is assumed that the order of gates does not change after the synthesis step. If two edges in the QODG come from one node and go to another node, the edges are combined in order to keep the graph *simple*. Also, due to the no-cloning theorem, a fan-out in the circuit is forbidden. A *start node* is added which connects to the first-level nodes in order to satisfy the initial dependencies. Also an *end node* is added where all the last-level nodes are connected to it. These two extra nodes simplify the problem formulation. A sample synthesized quantum circuit and the QODG constructed from it is presented in Figure 2.

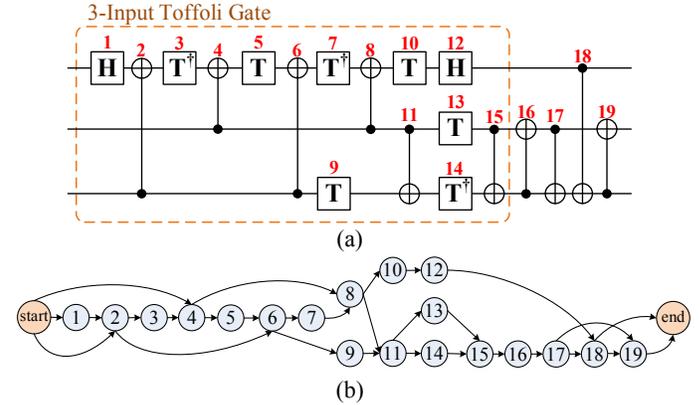

**Figure 2. (a) The synthesized *ham3* circuit [12] for size 3 Hamming optimal coding. Note that this circuit only contains FT gates. (b) A QODG constructed from the circuit shown in (a). Numbers are added to relate each node to its corresponding operation in the circuit.**

Based on the target quantum fabric and error threshold, a particular quantum coding is selected, and subsequently, a high-level tool maps the QODG into a TQA, where each ULB (tile) in this architecture can implement any operation in a fault-tolerant way. The latency of the quantum algorithm mapped to the TQA can be calculated as the length of the longest path (critical path) in the mapped QODG, where the length of a path in the QODG is the summation of latencies of operations located on the path plus routing latencies of their qubit operands. Note that the critical path of the mapped QODG may not be the same as the critical path of the original QODG because the latter does not contain routing latencies of logical qubits. These latencies change the scheduling slacks and hence may change the critical path of the entire graph.

Mapping a QODG to a TQA comprises of three intertwined steps: scheduling, placement, and routing. These steps depend on each other. For example, the result of placement and routing can increase the routing latency of a logical qubit and hence the qubit may fail to meet the timing requirements of the scheduling. As a result, the operation should be deferred by one or more scheduling steps. The quantum mapping problem, similar to the corresponding problem in the traditional VLSI area, is a hard problem. Hence, several heuristics have been proposed in the literature for solving it near-optimally [9][10][13][14]. Unfortunately, these heuristics are still very slow. They produce the mapping solution with the details of every qubit movement on the TQA. Since quantum computers are still not mature enough to handle large-scale problems [15], detailed information that a quantum mapper produces is excessive and not very useful. Hence, we introduce a model to quickly estimate the latency of a quantum algorithm as explained next.

# 3. ESTIMATING LATENCY OF A QUANTUM ALGORITHM

The latency of a quantum algorithm may be calculated as follows:

$$D = N_{CNOT}^{critical}(d_{CNOT} + L_{CNOT}^{avg}) + \sum_{g \in O} N_g^{critical}(d_g + L_g^{avg}) \quad (1)$$

where $O$ is the set of one-qubit FT operations (such as H, T, S, etc.); $N_{CNOT}^{critical}$ and $N_g^{critical}$ are the number of CNOTs and operations of type $g$ (one-qubit FT operations) on the critical path; $d_{CNOT}$ and $d_g$ determine the delay of CNOT and the operation of type $g$ respectively; $L_{CNOT}^{avg}$ and $L_g^{avg}$ capture the average routing latency for CNOT and the operation of type $g$. Note that the equation treats one and two-qubit operations differently. The only two-qubit FT operation is CNOT while there are different one-qubit operations. $N_{CNOT}^{critical}$ and $N_g^{critical}$ can be determined by calculating the critical path of the circuit. As explained earlier, considering the critical path of the original QODG instead of the critical path of the mapped QODG introduces some errors to the estimation model. So the values of $L_{CNOT}^{avg}$ and $L_g^{avg}$ will be added to the operation delays in the QODG in order to determine the critical path more accurately. $d_{CNOT}$ and $d_g$ depend on the underlying fabric technology, the error correction and the control techniques used. These parameters are the output of a ULB fabric designer tool which has a very low runtime execution (in the order of at most a few minutes) and produces exact results which can be used for any algorithms. Hence, values of these parameters for all types of FT operations are assumed to be given. $L_g^{avg}$ can be estimated empirically since routing of a qubit is not too complex. A one-qubit operation can be done in the ULB where the qubit currently resides or in the nearest free ULB if the current location is also occupied by another qubit. Value of $L_g^{avg}$ is set to $2 \times T_{move}$ where $T_{move}$ is a physical parameter which captures the time that a logical qubit needs to move from any ULBs, channels, or quantum crossbars to another ULB, channel or quantum crossbar in its neighborhood. This empirical result shows that on average each qubit needs to move to its nearest ULB for a one-qubit operation. The main challenge is to estimate $L_{CNOT}^{avg}$ which is more interesting as it represents the average traveling (routing) time of two logical qubits from their source locations to the target ULB (i.e., the ULB where the two qubits will interact). This value accounts for the traffic congestion in the routing channels. In this paper, a procedural method for estimating $L_{CNOT}^{avg}$ is suggested. Knowing this value and estimating the critical path, one can calculate the latency of a quantum program using Equation (1).

## 3.1 Estimating the Average Routing Latency for CNOT

The wire length estimation problem in the traditional VLSI area [16] approximates the average total wire length among all of the connected standard logic cells before performing the time-consuming cell placement and routing steps. Our problem is similar to the aforesaid, but in fact it is more complex. This is mainly because $L_{CNOT}^{avg}$ also depends on the scheduling of a QODG. More precisely, mapping of a QODG to a quantum fabric consists of three steps: scheduling, placement and routing. These steps are interrelated, and none can be optimally solved without solving the others. Placement and routing affect the result of scheduling (which in turn affects the timing slacks in the QODG.) Placement cannot be done optimally without considering the effect of routing and channel congestions. Also note that in the placement problem, one should assign both the logical qubits and the logical operations to ULBs. Compared to the VLSI placement, this problem has (*dynamically*) *moveable cells* since the qubits move during the execution of a program. Also two or more operations may be assigned to a ULB as long as they are scheduled to be done in different time slots. Moreover, the size of QODG for real-size problems is generally far larger than any standard VLSI gate-level netlists [10]. So, this estimation problem is more complex than the traditional VLSI counterpart.

For each logical qubit, a hypothetical presence zone is assumed in which the qubit performs most of its interactions. This zone also shows the area where the other qubits that interact with the qubit in question are located at some point in time. These zones are located in different places of the TQA fabric. They can overlap with each other. An overlap resembles congestion since it is possible that more than one qubit pass the overlapping area at the same time. Figure 3 depicts an illustration of five presence zones placed randomly on a fabric showing the interaction among five qubits. The overlapping area among zones 3, 4, and 5 is the most congested area.

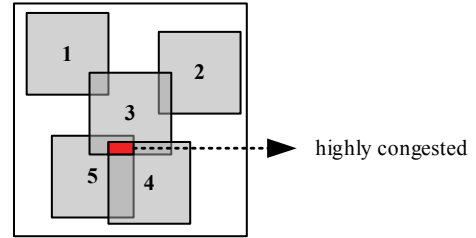

**Figure 3 Five presence zones placed randomly on the fabric**

Since the result of the placement is not known a priori, the zones are assumed to be placed randomly (uniformly and independently) on the fabric. $L_{CNOT}^{avg}$ can be estimated using Equation (2).

$$L_{CNOT}^{avg} \approx \frac{\sum_{q=1}^{Q} E[S_q] \times d_q}{\sum_{q=1}^{Q} E[S_q]} \quad (2)$$

$$\sum_{q=0}^{Q} E[S_q] = A \quad (3)$$

where $Q$ is the total number of logical qubits; $E[S_q]$ is the expected surface of the fabric covered by exactly $q$ overlapping presence zones; $d_q$ is the average routing latency of a qubit in an average-size presence zone when all the routing channels are occupied by $q$ qubits; and $A$ is the area of the fabric and it is equal to the total number of ULBs assuming that each ULB is a $1 \times 1$ square. Equation (3) shows a constraint on $E[S_q]$. Note that the summation index in the constraint starts at 0 instead of 1 since some parts of the fabric may not be covered by any presence zones. Since calculating the latency for the unoccupied surface is meaningless, $E[S_q]/\sum_{q=1}^{Q} E[S_q]$ is used in Equation (2) as the normalized value for $E[S_q]$. To calculate $E[S_q]$, Equation (4) is used:

$$E[S_q] = \binom{Q}{q} \sum_{x=1}^{a} \sum_{y=1}^{b} (P_{x,y})^q (1 - P_{x,y})^{Q-q} \quad (4)$$

where $P_{x,y}$ is the probability that the ULB at position (x,y) on the fabric is being covered by a qubit's presence zone randomly placed on the fabric; $a$ and $b$ are width and length of the fabric. (Remember that a fabric is modeled as a grid of $a \times b$ square-shape ULBs and $a \times b = A$.) The coefficient is the number of ways to choose $q$ presence zones from the total presence zones (i.e. $Q$ which equals to the total number of logical qubits). The two summations add the probability that each of the ULBs on the fabric is covered by exactly $q$ presence zones. The overall equation calculates the expected surface of the fabric covered by exactly $q$ presence zones. Calculating this equation $Q$ times and using it in Equation (2) is time consuming. Hence, only the first 20 terms are calculated in practice. Simulation results show that this choice does not dramatically affect the accuracy of the estimation while it substantially improves the runtime of LEQA.

Equation (5) calculates $P_{x,y}$. (In the nominator, two min{.} functions are multiplied. Note that they are written in two lines.)

$$P_{x,y} = \frac{\begin{pmatrix} \min\{x, a-x+1, \lceil\sqrt{B}\rceil, a-\lceil\sqrt{B}\rceil+1\} \times \\ \min\{y, b-y+1, \lceil\sqrt{B}\rceil, b-\lceil\sqrt{B}\rceil+1\} \end{pmatrix}}{(a-\lceil\sqrt{B}\rceil+1)(b-\lceil\sqrt{B}\rceil+1)} \quad (5)$$

where B is the average area of presence zones. Figure 4 depicts how Equation (5) is derived. The nominator counts the number of ways that placing a presence zone of size ($\lceil\sqrt{B}\rceil \times \lceil\sqrt{B}\rceil$) on a ($a \times b$) fabric covers the ULB located at position (x,y). Min{.} functions are used to account for the boundary situations. The denominator counts the number of ways a ($\lceil\sqrt{B}\rceil \times \lceil\sqrt{B}\rceil$) presence zone can be placed on a ($a \times b$) fabric.

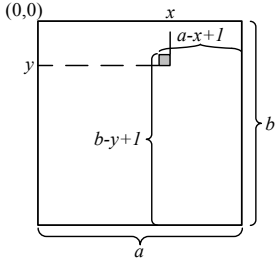

**Figure 4. Calculation of P$_{x,y}$**

To estimate $B$, the average area of presence zones, a new graph called *interaction intensity graph* IIG(V,E) is built as follows. Nodes of this graph are logical qubits which are denoted by $n_i$. An edge $e_{ij}$ is added between nodes $n_i$ and $n_j$ if these two qubits interact with each other. Weight of this edge, which is denoted by $w(e_{ij})$, is equal to the number of two-qubit operations between $n_i$ and $n_j$. Note that edges are not directed, so $e_{ij}$ and $e_{ji}$ refer to the same edge. Clearly, IIG(V,E) has no self-loops since no edges are added for one-qubit operations. $M_i$ is defined as the number of neighbors of node $n_i$ in the IIG(V,E). It is equal to $\deg(n_i)$ which is the degree of node $n_i$ in the IIG(V,E). We model the area of the presence zone associated with $n_i$, which is denoted by $B_i$, as follows:

$$B_i = \sqrt{M_i + 1} \times \sqrt{M_i + 1} \quad (6)$$

Addition of one to the term $M_i$ accounts for the qubit $n_i$ itself. (There are $M_i + 1$ qubits in the presence zone.) Qubit $n_i$ travels inside this zone and interacts with $M_i$ qubits. It visits $\sum_{\forall n_j \in \text{adj}(n_i)} w(e_{ij})$ number of ULBs which may not be necessarily unique during the program execution. The average size of presence zones, B, can be calculated by using a weighted average over the size of the presence zone of all logical qubits:

$$B = \frac{\sum_{i=1}^{Q}\left(\sum_{\forall n_j \in \text{adj}(n_i)} w(e_{ij})\right) \times B_i}{\sum_{i=1}^{Q} \sum_{\forall n_j \in \text{adj}(n_i)} w(e_{ij})} \quad (7)$$

$\sum_{\forall n_j \in \text{adj}(n_i)} w(e_{ij})$ sums over the weights of all adjacent edges of the node $n_i$ in IIG(V,E). It increases the weight of the term $B_i$ if the qubit $n_i$ is involved in more two-qubit operations.

To calculate $d_q$, which was used in Equation (2), the following equation is used:

$$d_q = \begin{cases} d_{uncong}, & q \leq N_c \\ \frac{(1+q)d_{uncong}}{N_c}, & \text{otherwise} \end{cases} \quad (8)$$

where $N_c$ is the capacity of routing channels and $d_{uncong}$ is the average routing latency of a qubit for interacting with another qubits in an average-size presence zone when all the routing channels are *uncongested*. A channel is considered as *uncongested* if the number of qubits inhibiting the channel is less than or equal to $N_c$. In this case, qubits can pass through the channel with the minimum delay (i.e. $d_{uncong}$). If $q$ is greater than $N_c$, the channel is called *congested* and qubits will form a pipeline for passing through it (hence, they will be delayed depending on their position in the pipeline). We capture this increase in the routing latency by modeling the routing channels as an M/M/1/∞ queue. Figure 5 shows a pictorial view of this model. Green blocks show logical qubits that are currently using the channel. Red blocks show the qubits waiting to get access to the channel. We assume that the arrival rate of qubits has Poisson distribution with parameter $\lambda$ since the inter-arrival time of qubits are independent and memory-less. Hence, a Poisson distribution can model it very well. The service rate is assumed to have an exponential distribution with parameter $\mu$. This assumption is made to simplify the calculations. Experimental results show that this simple model performs well in practice.

Knowing that the average routing latency for each qubit under the service is $d_{uncong}$, $\mu$ can be calculated as $N_c/d_{uncong}$. Moreover, the average length of the queue, $l_{queue}^{avg}$, is q which is the number of qubits in the queue.

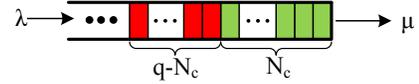

**Figure 5. An M/M/1/∞ queue model for routing channels**

Based on the queuing theory [17], $l_{queue}^{avg}$, (i.e., the average length of the queue) can be calculated as in Equation (9). Exploiting this equation and knowing the value of $l_{queue}^{avg}$, $\lambda$ can be calculated as shown in Equation (10).

$$l_{queue}^{avg} = \frac{\lambda}{\mu - \lambda} = \frac{\lambda}{\frac{N_c}{d_{uncong}} - \lambda} \quad (9)$$

$$q = \frac{\lambda}{\frac{N_c}{d_{uncong}} - \lambda} \rightarrow \lambda = \frac{qN_c}{(1+q)d_{uncong}} \quad (10)$$

Now the values of the arrival rate ($\lambda$) and the average queue length ($l_{queue}^{avg}$) are known. With these values, Little's formula [17] gives the average waiting (service) time in the queue ($W_{avg}$):

$$q = \frac{qN_c}{(1+q)d_{uncong}} \times W_{avg} \rightarrow W_{avg} = \frac{(1+q)d_{uncong}}{N_c} \quad (11)$$

This is the expression used in Equation (8). Estimation of $d_{uncong}$ is not a trivial task and explained in the next section.

### 3.2 Estimating $d_{uncong}$

To estimate $d_{uncong}$, a new parameter called $d_{uncong,i}$ is defined. This variable represents the average routing latency of the qubit $n_i$ in an average-size presence zone when the routing channels are not congested. A weighted average over all $d_{uncong,i}$ values, similar to the Equation (7), gives an estimation of $d_{uncong}$:

$$d_{uncong} = \frac{\sum_{i=1}^{Q}\left(\sum_{\forall n_j \in \text{adj}(n_i)} w(e_{ij})\right) \times d_{uncong,i}}{\sum_{i=1}^{Q} \sum_{\forall n_j \in \text{adj}(n_i)} w(e_{ij})} \quad (12)$$

One way to estimate $d_{uncong,i}$ is to randomly place $M_i + 1$ qubits in the presence zone of the qubit $n_i$ and calculate the expected length of the shortest Hamiltonian path ($E[l_{ham,i}]$) which goes through these qubits. These qubits can be placed anywhere in the presence zone, even they can be placed at the same location. This captures the fact that two qubit can travel to the same ULB for interaction. The reason for selecting Hamiltonian path is that according to the assumption, in a presence zone, only one qubit interacts with others, so it has to travel to $M_i$ locations (not necessarily unique) and interact with $M_i$ unique qubits. Interactions among other qubits are considered in their own presence zone calculation. A shortcoming of the aforementioned approach is that the problem of calculating the expected shortest Hamiltonian path is NP-hard [18]. Hence, the exact calculation of $E[l_{ham,i}]$ is infeasible for a *quick* estimation method. An upper bound and a lower bound for the expected path length of *traveling salesman problem* (TSP) are presented in reference [19]. It assumes that ($M_i + 1$) ≫ 1 points are randomly distributed in a $1 \times 1$ square. Equation

(13) presents a lower bound and equation (14) shows an upper bound for the expected path length of TSP.

$$\text{lower bound: } 0.708\sqrt{M_i + 1} + 0.551 \quad (13)$$

$$\text{upper bound: } 0.718\sqrt{M_i + 1} + 0.731 \quad (14)$$

Taking the average of the upper bound and the lower bound gives a good estimation for the expected path length of TSP. In our problem, the square length is $\sqrt{B_i}$ times greater so the result should be multiplied by $\sqrt{B_i}$. Moreover, since TSP solution is a tour, the result should also be multiplied by $(M_i - 1)/M_i$ to give the Hamiltonian path length which has one edge less than the tour. Equation (15) shows the resultant estimation for $E[l_{ham,i}]$.

$$E[l_{ham,i}] \approx \sqrt{B_i} \times (0.713\sqrt{M_i + 1} + 0.641) \times \frac{M_i - 1}{M_i} \quad (15)$$

By knowing the value of $E[l_{ham,i}]$, $d_{uncong,i}$ can be calculated as follows:

$$d_{uncong,i} = \frac{E[l_{ham,i}]}{v \times M_i} \quad (16)$$

where $v$ is a parameter depending on the physical characteristics of the fabric technology mostly the speed of moving a logical qubit through the channels. This parameter also can be used for tuning the LEQA with different quantum mappers. $M_i$ is presented in the denominator to give the average routing latency for an operation.

### 3.3 LEQA Algorithm and Its Performance

Algorithm 1 shows the implementation of LEQA based on the presented procedural method. Note that QODG is an input of the algorithm. One can easily construct it from a synthesized quantum circuit as shown in Figure 2. Size of the fabric is another input. This value can be changed to find the optimal size for the fabric which results in the minimum delay. The other inputs are physical parameters. The runtime complexity of the algorithm may be summarized as follows:

$$O(|V_{QODG}| + |E_{QODG}| + Q.A.\log Q) \quad (17)$$

More details on the analytical analysis to derive this time complexity are presented in the Supplemental Material section.

## 4. EXPERIMENTAL RESULTS

### 4.1 Simulation Setup

LEQA is implemented in Java. For the baseline, a quantum scheduling, placement, and routing tool (called *QSPR*) [20] was used. QSPR was minimally modified to work on the tile-based architecture of Figure 1. Table 1 lists the physical parameters of the TQA used for simulations. QSPR was also used to calculate the delay of performing FT operations on an ion-trap circuit fabric (left table). The [[7,1,3]] Steane code was used as the encoding and error correction scheme. Hence, delays of the T and T† gates ($d_T$ and $d_{T^\dagger}$) which are non-transversal in this coding, are higher than the others. These numbers can be adjusted based on any underlying technologies and does not limit the functionality of LEQA to a specific quantum realization technique. In the right table, the specifications of a TQA are presented.

**Table 1. List of physical parameters of the TQA**

| Parameter | Value | Parameter | Value |
|---|---|---|---|
| $d_H$ | 5440μs | $N_c$ | 5 |
| $d_T, d_{T^\dagger}$ | 10940μs | $v$ | 0.001 |
| $d_x, d_y, d_z$ | 5240μs | $A = a \times b$ | $3600 = 60 \times 60$ |
| $d_{CNOT}$ | 4930μs | $T_{move}$ | 100μs |

Benchmarks are taken from reference [12] and synthesized using the fault-tolerant gate library. The simple method presented in reference [4] is used to decompose n-input Toffoli and n-input Fredking gates (n>3) to several 3-input Toffoli and Fredking gates. Note that this method adds ancillary qubits to the circuit. Also no ancillary sharing is performed among the decomposed gates. The resultant 3-input Fredkin gates are replaced by three 3-input Toffoli gates. Finally, 3-input

**Algorithm 1: LEQA**

**Inputs:** QODG quantum operation dependency graph, $a, b$ width and length of the fabric, $d_{CNOT}$ and $d_g$ delays of logical gates, $N_c$ the capacity of routing channels, $v$ speed of a logical qubit through the routing channels, $Q$ number of logical qubits
**Outputs:** $D$ estimated latency of the input program

1. Make IIG(V,E) from the given QODG
2. Let $M_i = \deg(n_i)$ for every $n_i$ and calculate $B_i$ from Eq (6).
3. Calculate $B$ from Eq (7).
4. **For** ($i = 1$ to $Q$)
5.    Calculate $E[l_{ham,i}]$ using Eq (15).
6.    Calculate $d_{uncong,i}$ using Eq (16).
7. **End**
8. Calculate $d_{uncong}$ from Eq (12).
9. **For** (x= 1 to $a$)
10.    **For** (y= 1 to $b$)
11.      Calculate $P_{x,y}$ using Eq (5).
12.    **End**
13. **End**
14. **For** ($q = 1$ to $Q$)
15.    Calculate $d_q$ from Eq (8).
16.    Calculate $E[S_q]$ from Eq (4).
17. **End**
18. Calculate $L_{CNOT}^{avg}$ from the approximation given in Eq (2)
19. Update the $QODG$ based on the value of $L_{CNOT}^{avg}$ and empirical value for $L_g^{avg}$ and then calculate $N_{CNOT}^{critical}$ and $N_g^{critical}$ for all operations types
20. Calculate $D$ using the estimation given in Eq (1).
21. **Return** D

Toffoli gates are decomposed to a set of fault-tolerant gates using the method presented in reference [21] and shown in Figure 2.

LEQA and QSPR share the same parsers for parsing the inputs, the TQA specification, and physical parameters. A PC with Intel Pentium Dual-Core E5500 CPU clocked at 2.80GHz with 4GB RAM running Windows 7 and Java Development Kit (JDK) 7 is used for simulations.

### 4.2 Simulation Results

Table 2 shows the comparison between the actual delay computed by QSPR and the estimated delay calculated by LEQA. As can be seen, the average estimation error is equal to 2.11% while the maximum error is below 9%.

Table 3 lists the information about the benchmarks as well, i.e. the qubit count and operation count. The benchmarks are sorted based on the operation count. Also Table 3 compares the runtime of LEQA and QSPR. Evidently, when the operation count grows, LEQA performs better. In the largest benchmark, which its netlist file size is more than 12MB, LEQA performs more than two orders of magnitude faster than QSPR. This trend shows that as the size of netlist grows, LEQA beats QSPR in terms of speed and still gives accurate results.

As an interesting case, consider the last two benchmarks, i.e. *gf2^128mult* and *gf2^256mult*. The operation count of the latter benchmark is almost 4 times of the former one. By comparing the runtime of LEQA and QSPR for these two benchmarks, it can be seen that runtime of LEQA is increased by a factor of 3 while the runtime of QSPR is increased by a factor of 4.5. This further depicts the scalability of LEQA compared to QSPR.

LEQA achieves 114X speedup over QSPR for the largest benchmark, i.e. gf2^256mult. This factor increases for larger benchmarks. Precisely, QSPR runtime scales super linearly with operation count in the circuit (with degree of 1.5) whereas LEQA runtime depends only linearly on this count (see Equation (17)). Reference [10] reports that Shor algorithm for a 1024-bit integer has $1.35 \times 10^{15}$ physical operations. Using two-level [[7,1,3]] Steane code, each logical operation results in about $10^5$ physical operations. So this algorithm has almost $1.35 \times 10^{10}$ logical operations. Using extrapolation, QSPR would compute the latency in ~2 years whereas LEQA needs only 16.5 hours!! Moreover, multiple QSPR runs are needed to select minimum overhead QECC design.

**Table 2. Comparison between the actual latency computed by QSPR and the estimated latency calculated by LEQA**

| Benchmark | Actual Delay (sec) | Estimated Delay (sec) | Absolute Error (%) |
|---|---|---|---|
| 8bitadder | 1.617E+00 | 1.667E+00 | 3.10 |
| gf2^16mult | 4.460E+00 | 4.524E+00 | 1.45 |
| hwb15ps | 1.940E+01 | 1.993E+01 | 2.76 |
| hwb16ps | 1.852E+01 | 1.903E+01 | 2.76 |
| gf2^18mult | 5.085E+00 | 5.109E+00 | 0.46 |
| gf2^19mult | 5.393E+00 | 5.407E+00 | 0.25 |
| gf2^20mult | 5.654E+00 | 5.660E+00 | 0.11 |
| ham15 | 2.518E+01 | 2.530E+01 | 0.51 |
| hwb20ps | 3.026E+01 | 3.106E+01 | 2.66 |
| hwb50ps | 1.236E+02 | 1.274E+02 | 3.10 |
| gf2^50mult | 1.474E+01 | 1.495E+01 | 1.44 |
| mod1048576adder | 2.027E+02 | 1.958E+02 | 3.38 |
| gf2^64mult | 1.904E+01 | 1.935E+01 | 1.64 |
| hwb100ps | 3.427E+02 | 3.402E+02 | 0.72 |
| gf2^100mult | 3.015E+01 | 2.998E+01 | 0.57 |
| hwb200ps | 9.638E+02 | 8.839E+02 | 8.29 |
| gf2^128mult | 3.886E+01 | 3.838E+01 | 1.24 |
| gf2^256mult | 7.936E+01 | 7.654E+01 | 3.55 |

**Table 3. Information about benchmark circuits and comparison between the runtime of QSPR and LEQA**

| Benchmark | Qubit Count | Operation Count | QSPR Runtime (sec) | LEQA Runtime (sec) | Speedup (X) |
|---|---|---|---|---|---|
| 8bitadder | 24 | 822 | 0.9 | 0.115 | 8.2 |
| gf2^16mult | 48 | 3,885 | 3.0 | 0.289 | 10.3 |
| hwb15ps | 47 | 3,885 | 2.7 | 0.256 | 10.7 |
| hwb16ps | 55 | 3,811 | 2.9 | 0.250 | 11.5 |
| gf2^18mult | 54 | 4,911 | 3.5 | 0.276 | 12.6 |
| gf2^19mult | 57 | 5,469 | 3.7 | 0.259 | 14.2 |
| gf2^20mult | 60 | 6,019 | 5.1 | 0.301 | 17.1 |
| ham15 | 146 | 5,308 | 4.3 | 0.257 | 16.6 |
| hwb20ps | 83 | 6,395 | 3.8 | 0.272 | 13.9 |
| hwb50ps | 370 | 25,370 | 11.8 | 0.450 | 26.3 |
| gf2^50mult | 150 | 37,647 | 16.9 | 0.398 | 42.5 |
| mod1048576adder | 1,180 | 37,070 | 20.2 | 0.382 | 52.8 |
| gf2^64mult | 192 | 61,629 | 29.4 | 0.461 | 63.8 |
| hwb100ps | 1,106 | 67,735 | 26.7 | 0.575 | 46.4 |
| gf2^100mult | 300 | 150,297 | 65.2 | 0.859 | 76.0 |
| hwb200ps | 3,145 | 175,490 | 66.7 | 0.915 | 72.9 |
| gf2^128mult | 384 | 246,141 | 106.0 | 1.381 | 78.3 |
| gf2^256mult | 768 | 983,805 | 524.8 | 4.576 | 114.7 |

## 5. CONCLUSION

This paper presented *LEQA*—a fast latency estimation tool for evaluating the latency of a quantum algorithm mapped to a tiled quantum architecture. It uses a procedural method to calculate the latency of an algorithm based on computing the neighborhood population counts of qubits. Simulation results showed that in mid-size circuits, LEQA is two orders of magnitude faster than the modern quantum mapper that performs detailed scheduling, placement and routing of the quantum instructions and qubits in a quantum operation dependency graph to a quantum fabric. This speedup is expected to increase superlinearly as a function of circuit size (operation count). Moreover, LEQA could produce quick estimates of the circuit latency with sufficient accuracy i.e., an average of 2.11% error.

## 6. ACKNOWLEDGEMENT

This research was supported by the Intelligence Advanced Research Projects Activity (IARPA) via Department of Interior National Business Center contract number D11PC20165. The U.S. Government is authorized to reproduce and distribute reprints for Governmental purposes notwithstanding any copyright annotation thereon. The views and conclusions contained herein are those of the authors and should not be interpreted as necessarily representing the official policies or endorsements, either expressed or implied, of IARPA, DoI/NBC, or the U.S. Government.

# Supplemental Material

## 1. PERFORMANCE ANALYSIS OF LEQA

The number of nodes in a QODG is equal to the number of operations in the circuit plus two (because of the dummy *start* and *end* nodes) and designated as $|V_{QODG}|$. The number of edges in this graph is also shown by $|E_{QODG}|$. Knowing these parameters, the runtime complexity of each line (or set of lines) in Algorithm 1 can be calculated as follows:

**Line 1:** Making of graph IIG(V,E) needs a traversal of QODG which takes $\mathcal{O}(|V_{QODG}| + |E_{QODG}|)$.

**Line 2:** Calculation of $M_i$ and $B_i$ can be done in $\mathcal{O}(Q)$.

**Line 3:** Calculating the weights need to sum over all edges in the IIG(V,E) which has at most $\mathcal{O}(|V_{QODG}|)$ edges. Calculating the summation over weighted $B_i$s takes $\mathcal{O}(Q)$. Overall this line takes $\mathcal{O}(|V_{QODG}| + Q)$ to be done.

**Lines 4-7:** Calculation of $E[l_{ham,i}]$ and $d_{uncong,i}$ can be done in constant time and hence the for-loop takes $\mathcal{O}(Q)$ to complete.

**Line 8:** Same as line 3, it takes $\mathcal{O}(|V_{QODG}| + Q)$. One can reuse the calculated weights in line 3 to reduce the calculation time to $\mathcal{O}(Q)$.

**Lines 9-13:** The nested for-loops iterate $A(= a \times b)$ times in total. In each iteration, the value of $P_{ab}$ is calculated in constant time. So it takes $\mathcal{O}(A)$ time to complete.

**Lines 14-17:** The for-loop iterates $Q$ times and in each iteration, line 15 takes $\mathcal{O}(1)$ whereas line 16 takes $\mathcal{O}(A.\log Q)$. $\mathcal{O}(A)$ is the result of the double summation over the area and $\mathcal{O}(\log Q)$ is the time needed to calculate $(P_{x,y})^q$ and $(1 - P_{x,y})^{Q-q}$. The value of $\binom{Q}{q}$ can be calculated in constant time using the following recursive formula:

$$f(Q, 0) = 1$$
$$f(Q, q) = f(Q, q-1) \times \frac{Q - q + 1}{q}, \quad 0 < q \leq Q \quad (18)$$

Overall these lines take $\mathcal{O}(Q.A.\log Q)$ for completion. As explained in the paper, only the first 20 values for $E[S_q]$ is calculated in practice, i.e. for $q = 1$ to 20. Hence, in action LEQA performs much faster than $\mathcal{O}(Q.A.\log Q)$.

**Line 18:** The calculation takes $\mathcal{O}(Q)$.

**Line 19:** Updating the delay of all instructions takes $\mathcal{O}(|V_{QODG}|)$. Calculation of the critical path in a directed acyclic graph (DAG) takes $\mathcal{O}(|V_{QODG}| + |E_{QODG}|)$ (Chapter 24 of the reference [1] explains an algorithm with this time complexity). Deriving the values of $N_g^{critical}$ and $N_{CNOT}^{critical}$ can be done by traversing the critical path which has the length $\mathcal{O}(|V_{QODG}|)$ in the worst case.

**Line 20:** Calculation of $D$ can be done in constant time.

So, the overall runtime of the algorithm may be summarized as follows:

$$\mathcal{O}(|V_{QODG}| + |E_{QODG}| + Q.A.\log Q) \quad (19)$$

## 2. REFERENCE

[1] T. H. Cormen, C. E. Leiserson, R. L. Rivest, and C. Stein, *Introduction to Algorithms*, 3rd ed. The MIT Press, 2009.